\documentclass[aps,prl,preprint,tightenlines,superscriptaddress,showpacs,byrevtex]{revtex4}
\usepackage{graphicx} 
\usepackage{dcolumn}  

\graphicspath{{ps}}

\begin{document}


\preprint{\vbox{ \hbox{   }
                 \hbox{BELLE-CONF-0533}
                 \hbox{LP2005-168}
                 \hbox{EPS05-508} 
}}

\title{ \quad\\[0.5cm]  Study of $B^0 \to \eta K^+ \pi^-$ and $\eta \pi^+ \pi^-$}
\affiliation{Aomori University, Aomori}
\affiliation{Budker Institute of Nuclear Physics, Novosibirsk}
\affiliation{Chiba University, Chiba}
\affiliation{Chonnam National University, Kwangju}
\affiliation{University of Cincinnati, Cincinnati, Ohio 45221}
\affiliation{University of Frankfurt, Frankfurt}
\affiliation{Gyeongsang National University, Chinju}
\affiliation{University of Hawaii, Honolulu, Hawaii 96822}
\affiliation{High Energy Accelerator Research Organization (KEK), Tsukuba}
\affiliation{Hiroshima Institute of Technology, Hiroshima}
\affiliation{Institute of High Energy Physics, Chinese Academy of Sciences, Beijing}
\affiliation{Institute of High Energy Physics, Vienna}
\affiliation{Institute for Theoretical and Experimental Physics, Moscow}
\affiliation{J. Stefan Institute, Ljubljana}
\affiliation{Kanagawa University, Yokohama}
\affiliation{Korea University, Seoul}
\affiliation{Kyoto University, Kyoto}
\affiliation{Kyungpook National University, Taegu}
\affiliation{Swiss Federal Institute of Technology of Lausanne, EPFL, Lausanne}
\affiliation{University of Ljubljana, Ljubljana}
\affiliation{University of Maribor, Maribor}
\affiliation{University of Melbourne, Victoria}
\affiliation{Nagoya University, Nagoya}
\affiliation{Nara Women's University, Nara}
\affiliation{National Central University, Chung-li}
\affiliation{National Kaohsiung Normal University, Kaohsiung}
\affiliation{National United University, Miao Li}
\affiliation{Department of Physics, National Taiwan University, Taipei}
\affiliation{H. Niewodniczanski Institute of Nuclear Physics, Krakow}
\affiliation{Nippon Dental University, Niigata}
\affiliation{Niigata University, Niigata}
\affiliation{Nova Gorica Polytechnic, Nova Gorica}
\affiliation{Osaka City University, Osaka}
\affiliation{Osaka University, Osaka}
\affiliation{Panjab University, Chandigarh}
\affiliation{Peking University, Beijing}
\affiliation{Princeton University, Princeton, New Jersey 08544}
\affiliation{RIKEN BNL Research Center, Upton, New York 11973}
\affiliation{Saga University, Saga}
\affiliation{University of Science and Technology of China, Hefei}
\affiliation{Seoul National University, Seoul}
\affiliation{Shinshu University, Nagano}
\affiliation{Sungkyunkwan University, Suwon}
\affiliation{University of Sydney, Sydney NSW}
\affiliation{Tata Institute of Fundamental Research, Bombay}
\affiliation{Toho University, Funabashi}
\affiliation{Tohoku Gakuin University, Tagajo}
\affiliation{Tohoku University, Sendai}
\affiliation{Department of Physics, University of Tokyo, Tokyo}
\affiliation{Tokyo Institute of Technology, Tokyo}
\affiliation{Tokyo Metropolitan University, Tokyo}
\affiliation{Tokyo University of Agriculture and Technology, Tokyo}
\affiliation{Toyama National College of Maritime Technology, Toyama}
\affiliation{University of Tsukuba, Tsukuba}
\affiliation{Utkal University, Bhubaneswer}
\affiliation{Virginia Polytechnic Institute and State University, Blacksburg, Virginia 24061}
\affiliation{Yonsei University, Seoul}
  \author{K.~Abe}\affiliation{High Energy Accelerator Research Organization (KEK), Tsukuba} 
  \author{K.~Abe}\affiliation{Tohoku Gakuin University, Tagajo} 
  \author{I.~Adachi}\affiliation{High Energy Accelerator Research Organization (KEK), Tsukuba} 
  \author{H.~Aihara}\affiliation{Department of Physics, University of Tokyo, Tokyo} 
  \author{K.~Aoki}\affiliation{Nagoya University, Nagoya} 
  \author{K.~Arinstein}\affiliation{Budker Institute of Nuclear Physics, Novosibirsk} 
  \author{Y.~Asano}\affiliation{University of Tsukuba, Tsukuba} 
  \author{T.~Aso}\affiliation{Toyama National College of Maritime Technology, Toyama} 
  \author{V.~Aulchenko}\affiliation{Budker Institute of Nuclear Physics, Novosibirsk} 
  \author{T.~Aushev}\affiliation{Institute for Theoretical and Experimental Physics, Moscow} 
% ...
\collaboration{The Belle Collaboration}

\collaboration{Belle Collaboration}
\noaffiliation
\begin{abstract}
We report results of studies of
inclusive $B \to \eta K^+ \pi^-$ and $B \to \eta \pi^+ \pi^-$ decays. Charged conjugates are implied throughout this paper. These are obtained from a  data sample containing 386
million $B\bar{B}$ pairs, collected
at the $\Upsilon(4S)$ resonance,
with the Belle detector at the KEKB asymmetric energy $e^+ e^-$
collider.
The branching fraction of inclusive $B^0 \to \eta K^+\pi^-$ and $B^0 \to \eta \pi^+\pi^-$ are measured to be
$\mathcal B$($B^0 \to \eta K^+ \pi^-$)=
$(31.7\pm1.9 ^{+2.2}_{-2.6})\times 10^{-6}$ and $\mathcal B$($B^0 \to \eta \pi^+ \pi^-$)=$(6.2^{ +1.8 + 0.8}_{ -1.6 - 0.6})\times 10^{-6}$, 
where
the first error is statistical and the second systematic.
The decays $B^0\to a_0^- X^+$, where $X^+$= $K^+$, $\pi^+$ were searched for and no significant signals found.
Upper limits of $\mathcal B$($B^0 \to a_0^- K^+$)$<1.6 \times 10^{-6}$ and $\mathcal B$($B^0 \to a_0^- \pi^+$)$< 2.8 \times 10^{-6}$ at 90\% C.L. are obtained. Here the notation $B(B^0 \to a_0^- X^+)$ indicates the
product of branching fractions for $B^0 \to a_0^- X^+$ and 
$a_0^- \to \eta \pi^-$.
\end{abstract}
\maketitle
\tighten
{\renewcommand{\thefootnote}{\fnsymbol{footnote}}}
\setcounter{footnote}{0}
\section{Introduction} 
Recently, observations of large branching fractions of B mesons to three-body charmless hadronic systems 
have been reported by the $B$ factory experiments
 \cite{3body1}-\cite{3body4}. In the mesonic decays $B^+ \to K^+ \pi^+ \pi^-$ and 
$B^+ \to K^+ K^+K^-$, the broad $K^+K^-$ mass spectrum above 1.5 GeV/$c^2$ in 
$B^+ \to K^+ K^- K^+$ suggests a large non-resonant $B^+ \to K^+ K^- K^+$ 
contribution. In the baryonic decay $B^+ \to p\overline{p}K^+$, 
the $p\overline{p}$ mass spectrum cannot be explained by a simple phase-space
model. A baryonic form factor model\cite{baryon} or an additional
unknown resonance around 2 GeV/c$^2$ are both possible
explanations.
These studies of three-body decays have already provided new information on the 
mechanism of $B$ meson decay. Further, they suggest opportunities to search for previously  
unknown decays. Here we report results on $B$ meson decays to $\eta K^+ \pi^-$ and $\eta \pi^+ \pi^-$
 based on a data sample that
contains 386 million $B\overline{B}$ pairs,
collected  with the Belle detector at the KEKB asymmetric-energy
$e^+e^-$ (3.5~GeV on 8~GeV) collider~\cite{kekb}.
KEKB operates at the $\Upsilon(4S)$ resonance
($\sqrt{s}=10.58$~GeV) with a peak luminosity that has exceeded
$1.5\times 10^{34}~{\rm cm}^{-2}{\rm s}^{-1}$.

 There are several interesting quasi-two-body decays such as $B^0 \to a_0^- K^+$ and $B^0 \to a_0^- \pi^+$ included in this study. The observation of these modes will provide information both on $B$ meson decays to scalar mesons and on the nature of the $a_0^-$.

\section{Apparatus and Data Set}  
The Belle detector is a large-solid-angle magnetic
spectrometer that
consists of a silicon vertex detector (SVD),
a 50-layer central drift chamber (CDC), an array of
aerogel threshold \v{C}erenkov counters (ACC),
a barrel-like arrangement of time-of-flight
scintillation counters (TOF), and an electromagnetic calorimeter
comprised of CsI(Tl) crystals (ECL) located inside
a super-conducting solenoid coil that provides a 1.5~T
magnetic field.  An iron flux-return located outside of
the coil is instrumented to detect $K_L^0$ mesons and to identify
muons (KLM).  The detector
is described in detail elsewhere~\cite{Belle}.
Two inner detector configurations were used. A 2.0 cm beampipe
and a 3-layer silicon vertex detector was used for the first sample
of 152 million $B\bar{B}$ pairs (Set 1), while a 1.5 cm beampipe, a 4-layer
silicon detector and a small-cell inner drift chamber were used to record
the remaining 234 million $B\bar{B}$ pairs (Set II)\cite{Ushiroda}.

For Monte Carlo (MC) studies, samples of signal, generic b$\to$c decays and charmless rare $B$ decays are generated with the EVTGEN\cite{Evt} event generator. 
Continuum MC
events from the process $e^+e^- \to \gamma^* \to q \bar{q}$ are generated with
the JETSET\cite{JET} generator. 
The GEANT3~\cite{geant} package is used for detector simulation.
Signal features are studied with one hundred thousand MC events of
each decay mode. 
Background $B$ decays are studied with 706 million generic $B\bar{B}$ events and ``rare $B$'' MC corresponding to 21 times luminosity, where the latter are charmless hadronic and radiative B decays with
branching fractions
taken from the Particle Data Group~\cite{pdg}.  
 

\section{Event Selection and Reconstruction}

Hadronic events are selected based on the charged track multiplicity
and the total visible energy sum, which gives an efficiency
greater than 99\% for generic $B\bar B$ events.
All primary charged tracks are required to satisfy track quality cuts  
based on their impact parameters relative to the run-dependent  
interaction point (IP). The deviation from the IP is required to be within $\pm$ 0.1 cm
in the transverse direction and $\pm$ 2 cm in the longitudinal direction,
where the z (longitudinal) axis is taken as opposite to the positron beam direction. 
Due to the detector response, a polar angle cut ( $-0.866$ $<$ $ \cos \theta$ $<$ 0.956 ) is applied to all charged tracks. Only tracks not identified as muons or electrons are used in this analysis.
Particle identification (PID) is based on
${\mathcal L_{\rm K}}$/(${\mathcal L_\pi+L_{\rm K}}$)  
information, where ${\mathcal L}_{{\rm K}(\pi)}$ stands for the likelihood
for charged kaons (pions).  
PID cuts are applied to all the charged particles in this analysis. 
Unless otherwise stated, these are 
${\mathcal L_{\rm K}}$/(${\mathcal L_\pi+L_{\rm K}}$) $> 0.6$ for kaons
and ${\mathcal L_{\rm K}}$/(${\mathcal L_\pi+L_{\rm K}}$) $< 0.4$ for pions. The PID efficiencies are $85$\% for
kaons and $89$\% for pions, while the 
fake rates are $8$\% for pions faking kaons and 
$11$\% for kaons faking pions, respectively. 
Photon energies are required to be greater 
than 50~MeV (100~MeV) within the acceptance of the barrel (endcap) ECL. 
 
Candidate $\eta$ mesons are reconstructed through
$\eta \to \gamma \gamma$ and $\eta \to \pi^+\pi^- \pi^0$.
For $B$ reconstruction, the momenta of $\eta$ candidates are 
recalculated by applying the $\eta$ mass constraint in a vertex-mass constrained fit.
For $\eta \to \gamma\gamma$ decays, candidate $\eta$'s
are selected with $|\cos\theta^*| <0.90$, where $\theta^*$ 
is the angle between the photon direction in the $\eta$ rest frame 
and the $\eta$ momentum in the lab frame, to suppress the soft photon combinatorial
background and $B \to K^* \gamma$ feed-across. 
The $\eta$ mass regions are 0.500~GeV/$c^2 \,$-$\,$0.575~GeV/$c^2$ for  
$\eta \to \gamma \gamma$, and 0.535~GeV/$c^2\,$-$\,$0.560~GeV/$c^2$ for $\eta \to \pi^+ \pi^- \pi^0$.  
 
$K^{*0}$ mesons are reconstructed from $K^{*0} \to K^+ \pi^-$. 
Candidate $a_0^-$ mesons are reconstructed from $a_0^- \to \eta \pi^-$ and are required to have masses 
within 150 MeV/c$^2$ of the nominal value. For $a_0^- \to \eta \pi^-$, the modulus of the helicity($|\cos\theta_{{\rm{hel}}(a_0^-)}|$) must be less than 0.8, where $\theta_{{\rm{hel}}(a_0^-)}$ is the angle between the $\pi^+$ and $B^0$ in the $a_0^-$ rest frame.  

Note that, in the inclusive $B^0 \to \eta K^+ \pi^-$ channel, we have applied a $\overline{D}{}^0 \to K^+ \pi^-$(1.841 GeV/c$^2$ $<$ $M_{K\pi}$ $<$ 1.888 GeV/c$^2$) veto
and in the inclusive $B^0 \to \eta \pi^+ \pi^-$ channel, we have applied $D^0 \to \pi^+
\pi^-$(1.80 GeV/c$^2$ $<$ $M_{\pi\pi}$ $<$ 1.94 GeV/c$^2$) and $D^- \to \eta \pi^-$(1.80 GeV/c$^2$ $<$ $M_{\eta\pi}$ $<$ 1.94 GeV/c$^2$) vetos.
  
$B$ meson candidates are identified using the beam-energy constrained mass 
$M_{\rm{bc}}$ = $\sqrt{E^2_{\rm beam}-|P_B|^2}$ and the energy difference 
$\Delta E = E_B - E_{\rm beam}$, where $E_{\rm beam} = 5.29$~GeV, and 
($P_B$, $E_B$) is the four-momentum of the B candidate in the  
$\Upsilon(4S)$ rest frame. We impose a 2-D box cut
on $M_{\rm{bc}}$ and $\Delta E$ with
$M_{\rm{bc}} >$ 5.2~GeV/$c^2$
and $|\Delta E| <$ 0.3~GeV for further analysis. 
A signal region with $M_{\rm{bc}} > 5.27$~GeV/$c^2$ 
and $-0.1$~GeV$< \Delta E < 0.08$~GeV
is selected to plot the projections of fits.
A sideband region is defined with 
$M_{\rm{bc}} \le 5.26$~GeV/c$^2$ inside the box region.

At most one candidate per event in each mode is required. The best candidate 
is chosen based on the sum of $\chi^2$ of the $\eta$ vertex-mass constrained fit 
and the $\chi^2$ of the $K\pi(\pi\pi)$ vertex fit.

The dominant background for the three body $B$ decay events comes from $e^+e^- \to q\overline{q}$ continuum events, where $q$ = $u$,$d$,$s$ or $c$. In order to reduce this background, several shape variables are chosen to distinguish spherical $B\overline{B}$ events from jet-like continuum events. Five modified Fox-Wolfram moments \cite{fox} and a measure of the momentum transverse to the event thrust axis ($S_{\perp}$)~\cite{CLEO2} are combined into a Fisher discriminant~\cite{fisher}. The Probability Density Functions (PDFs) for this discriminant and $\cos \theta_{B}$, where $\theta_{B}$ is the angle between the $B$ flight direction and the beam direction in the $\Upsilon$(4$S$) rest frame, are obtained using events in the signal and sideband regions from MC simulations of signal and $q\overline{q}$ events. 
The displacement along the beam direction between the signal $B$ vertex and 
that of the other $B$, $\Delta z$, also provides separation. For $B$ events, the
 average value of $\Delta z$ is approximately 200 $\mu$m, while continuum events
 have a common vertex. Additional discrimination is provided by the $b$-flavor 
tagging algorithm~\cite{btag} developed for time-dependent analysis at Belle. 
The flavor tagging procedure yields two outputs: $q$ (= $\pm$ 1), which indicates the flavor of the tagging $B$,
and $r$, which ranges from 0 to 1, and is a measure of the confidence of the tag.
Events with high values of $r$ are well-tagged and are less likely to originate
 from continuum production. Thus, the quantity $q\cdot r$ can be used to discriminate against continuum events. 
The PDFs derived from the Fisher discriminant, the $\cos \theta_{B}$ distributions and the $\Delta z$ distributions are multiplied 
to form a likelihood ratio $\mathcal R$= $\mathcal L_s$/($\mathcal L_s+\mathcal L_{q\overline{q}}$), where $\mathcal L_s(L_{q\overline{q}})$ is the product of the 
signal($q\overline{q}$) probability densities. We achieve continuum background suppression by imposing $q\cdot r$-dependent $\mathcal R$ requirements, based on a study of the signal significance ($N_S/\sqrt{N_S+N_B}$) using a MC sample, where $N_S$ and $N_B$ are signal and background yields, respectively. 
The effect of the $\mathcal R$ cut is studied by comparing the cut efficiency on reconstructed $B^+ \to \overline{D}{}^0 \pi^+$ events in data and MC, for different values of $\mathcal R$. A systematic error of $\sim$2\% is obtained for the $\mathcal R$ cut.

The resolution of the signal $M_{\rm{bc}}$ width ($\sigma_{M_{\rm{bc}}}$) is verified using data and MC samples of reconstructed $B^+ \to \eta' K^+$, $\eta' \to \eta \pi^+ \pi^-$ events. Our MC underestimates $\sigma_{M_{\rm{bc}}}$ by 8.98$^{+5.76}_{-5.50}\,\%$ and 7.56$^{+6.96}_{-6.49}\,\%$ for the set I and set II data samples, respectively. An inclusive $\eta'$ sample, where $\eta' \to \eta \pi^+ \pi^-$, is used to check the $\Delta E$ resolution. The ratio of data to MC $\Delta E$ widths in the $\eta \to \gamma \gamma$ and $\eta \to \pi^+ \pi^- \pi^0$ modes are 1.03 $\pm$ 0.02 (1.05 $\pm$ 0.05) and 1.07 $\pm$ 0.03 (1.14 $\pm$ 0.05) for the set I(set II) data sample.

\section{Analysis Procedure}

Signal yields are obtained using an extended unbinned
maximum likelihood (2-D ML) fit to the $M_{\rm{bc}}$ and $\Delta E$
distributions in the $M_{\rm{bc}}-\Delta E$
box region after the ${\mathcal R}$
cut is applied.

For $N$ input candidates, the likelihood is defined as
\begin{eqnarray}
L(N_S,N_C) = \frac{e^{-(N_S+N_C+N_{b\bar b}+N_{ra})}}{N!} \prod_{i=1}^{N}
[N_{S} P_{S_i} + N_{C} P_{C_i} + N_{b\bar b} P_{{b\bar b}_i}
+N_{ra} P_{{ra}_i}],
\end{eqnarray}
where $P_{S_i}$, $P_{C_i}$, $P_{b{\bar b}_i}$ 
and $P_{{ra}_i}$ are the two-dimensional probability densities for
event $i$ to be the signal, continuum, charm containing ($b \to c$) $B$ decay backgrounds and
charmless $B$ decay backgrounds in the variables $M_{\rm{bc}}$
and $\Delta E$, respectively. Poisson fluctuations
for $N_S$ and $N_C$
are considered 
in this type of likelihood. 
For backgrounds other than continuum, $N_{b\bar b}$ and $N_{ra}$ 
are obtained from
MC samples. Uncertainties in the MC PDFs are included in the 
systematics study.
The continuum, $b \to c$ and charmless $B$ decay background PDFs are  
all obtained from the respective MC samples.

The $M_{\rm{bc}}$ and $\Delta E$ shapes from continuum MC events are modeled by an ARGUS function~\cite{argus} with a fixed end point at 5.29 GeV/c$^2$ and by a 2nd order Chebyshev polynomial, respectively. The shapes of signal, $B\overline{B}$, and other rare charmless $B$ decays are modeled by 2D smooth functions.
The $\Delta E$ distribution is found to be asymmetric, with a tail on the lower side due to $\gamma$ interactions with material in the front of the calorimeter and shower leakage out of the back side of the crystals. As a result, the $\Delta E$ resolution and the tail distribution strongly depend on the $\eta$ energy. In the inclusive $B^0 \to \eta K^+ \pi^-$ and $B^0 \to \eta \pi^+ \pi^-$ studies, the $\eta $ energy distribution for the signal events is not known apriori so we divide the data into three samples: $P_{\eta}<$ 1 GeV/$c$, 1 GeV/$c$ $<$ $P_{\eta}$ $<$ 2 GeV/$c$ and $P_{\eta} >$ 2 GeV/$c$.

For decays with more than one sub-decay process,
the final average results are obtained
by fitting the sub-decay modes simultaneously  
with the expected 
efficiencies included in the fit and with the branching fraction
as the common output.
This is equivalent to minimizing the sum of 
$\chi^2 = -2\ln(L)$
as a function of the branching fraction over all considered 
sub-decay channels. The statistical significance ($\Sigma$) of the signal
is
defined as $\sqrt{-2\ln(L_0/L_{\rm max})}$,
where $L_0$ and $L_{\rm max}$ denote
the likelihood values for zero signal events,
and the fit number of signal events, respectively.
The 90\% C.L. upper limit is calculated by finding $x_{\rm max}$ such that
\begin{eqnarray}
\frac{\int_0^{x_{\rm max}} L(x) \,dx}{\int_0^\infty L(x) \,dx}
 = 90\% \,.
\end{eqnarray}

\section{Measurement of Branching Fractions}

We correct our signal MC efficiency for several observed differences between it and data. Differences in the PID efficiencies are corrected using a 
$D^{*+} \to D^0 \pi^+$,
$D^0 \to K^- \pi^+$ control sample.
A study using $D^*$ partial reconstruction is used to correct the tracking efficiency.
To correct the $\pi^0$
reconstruction efficiency, a high momentum
inclusive $\eta$ sample is used where the ratios of reconstruced  
$\eta \to \pi^0 \pi^0 \pi^0$ to reconstruced $\eta \to \gamma\gamma$ in
 data and MC are compared.
The MC simulation for low energy photons is further tested by comparing
the $\eta$ helicity distribution for data with MC predictions.
The ratio of the single $\pi^0$ reconstruction efficiencies in data as compared to MC is 0.924 with a conservative systematic error of 3\%.
The ${\mathcal R}$ cut efficiency correction is determined
using $B^+ \to \overline{D}{}^0 \pi^+$ decays.
The high-momentum $\eta$ sample study is also used to correct the
efficiency of $\eta$ reconstruction and mass cuts. The correction factor for the mass cut of $\eta \to \gamma \gamma$($\pi\pi\pi^0$) is 0.990 $\pm$ 0.001 (0.993 $\pm$ 0.003)
All examined efficiencies show fairly good agreement
between data and MC samples.
The tracking, PID, $\pi^0$ and $\eta$ 
reconstruction efficiency systematic uncertainties are also obtained from the above studies.

%
\subsection{Inclusive $B^0 \to \eta K^+ \pi^-$ and $B^0 \to \eta \pi^+ \pi^-$}
\begin{figure}[htb]
\includegraphics[width=0.49\textwidth]{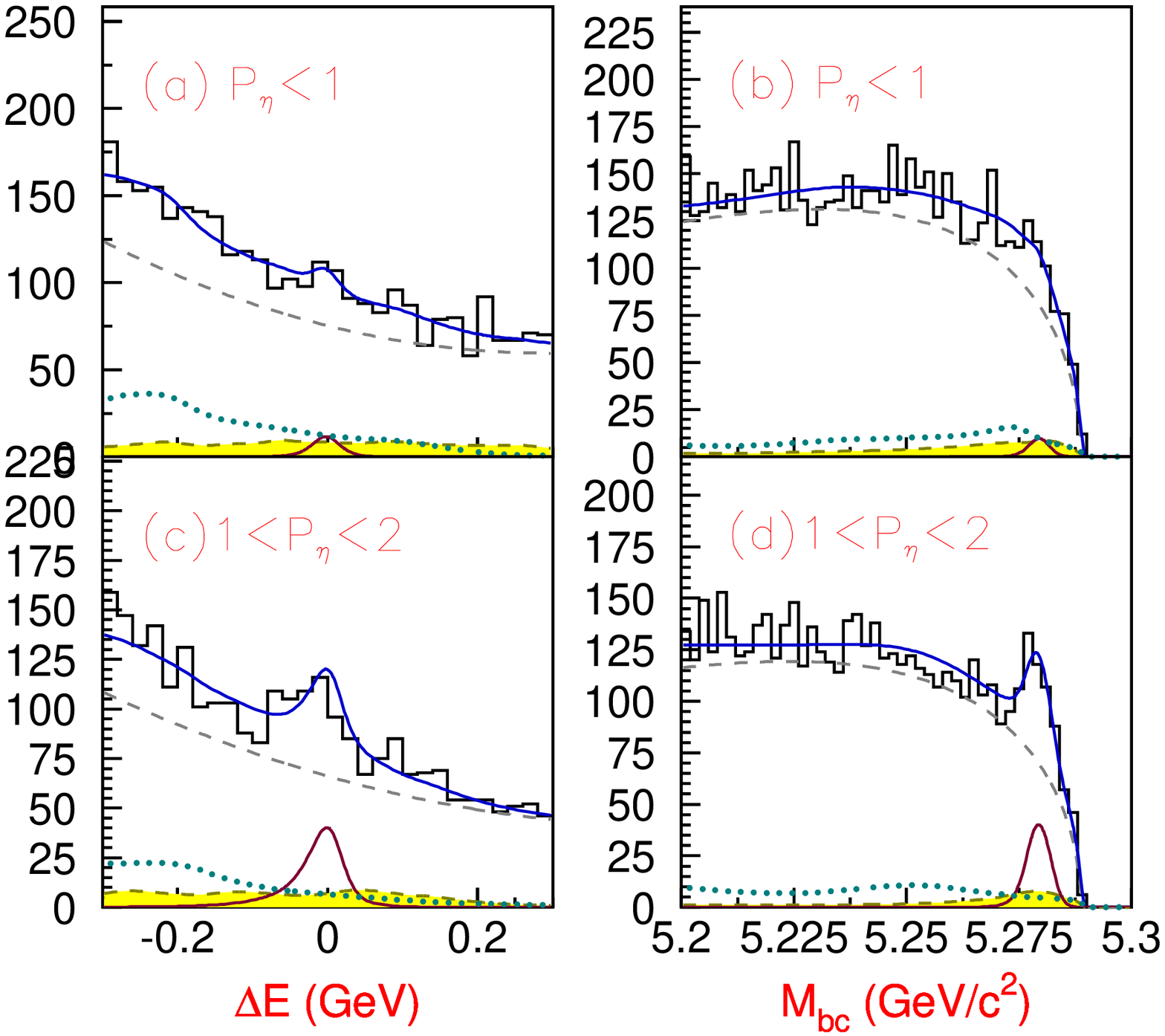}
\includegraphics[width=0.49\textwidth]{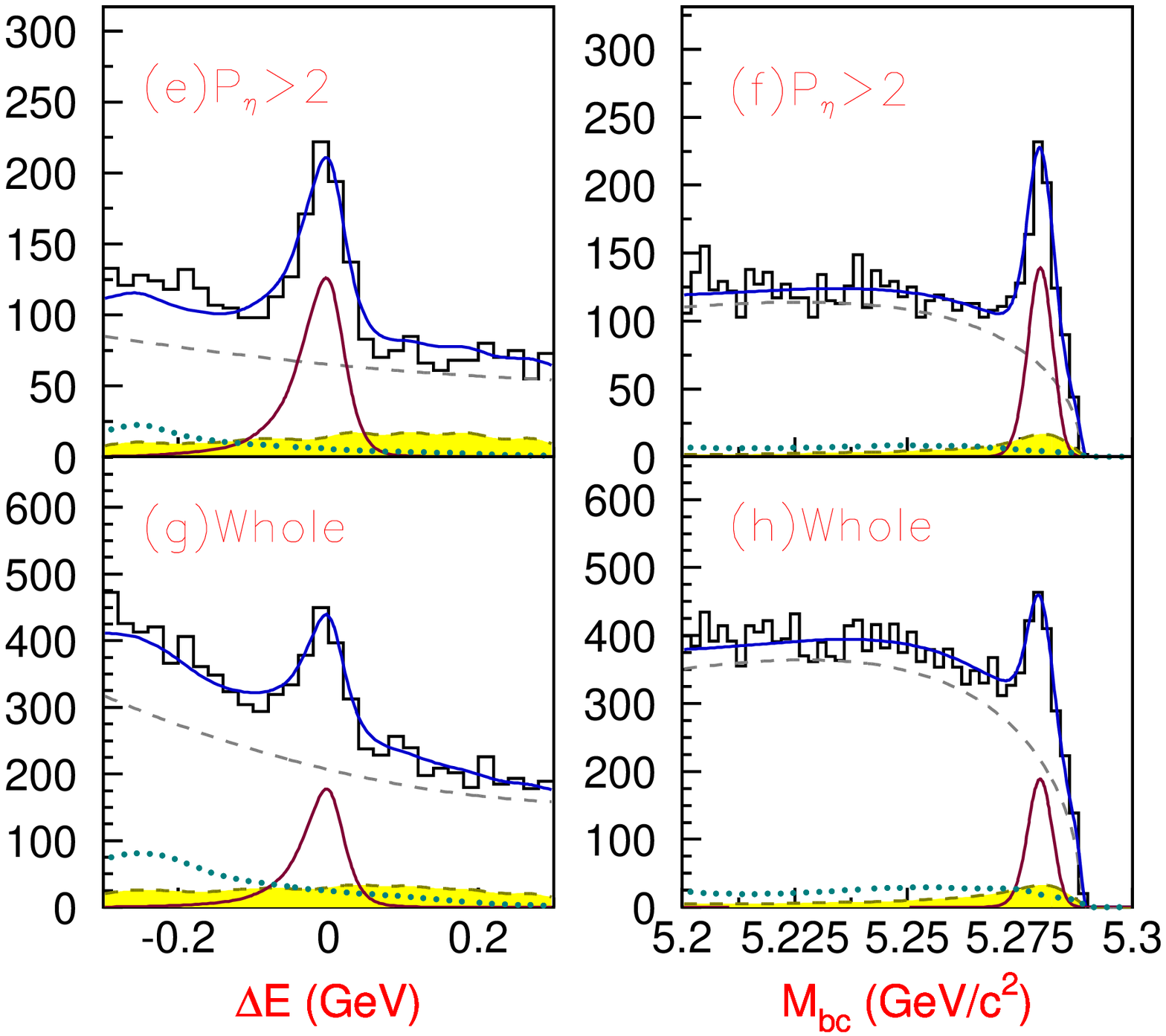}
\caption{\label{fitetakpi}Projections onto $M_{\rm{bc}}$ and $\Delta E$ from the extended unbinned 2D ML fit
for inclusive $B^0 \to \eta K^+ \pi^-$. The fit results with different $\eta$ momenta ranges are shown in the different plots. (a),(b) : $P_{\eta}<$ 1 GeV/c, (c),(d) :
1 GeV/c $<$ $P_{\eta}$ $<$ 2 GeV/c, (e),(f): $P_{\eta}>$ 2 GeV/c, (g),(h) : whole $P_\eta$ region. Gray dashed lines show the continuum $q\overline{q}$ contribution; Turquoise dotted lines show the generic $B\overline{B}$ background, yellow shaded parts show the other rare $B$ events; the red curves show the signal component; the blue curves show the sum of all above contributions; and the histograms show the data distribution.}
\end{figure}

\begin{figure}[htb]
\includegraphics[width=0.49\textwidth]{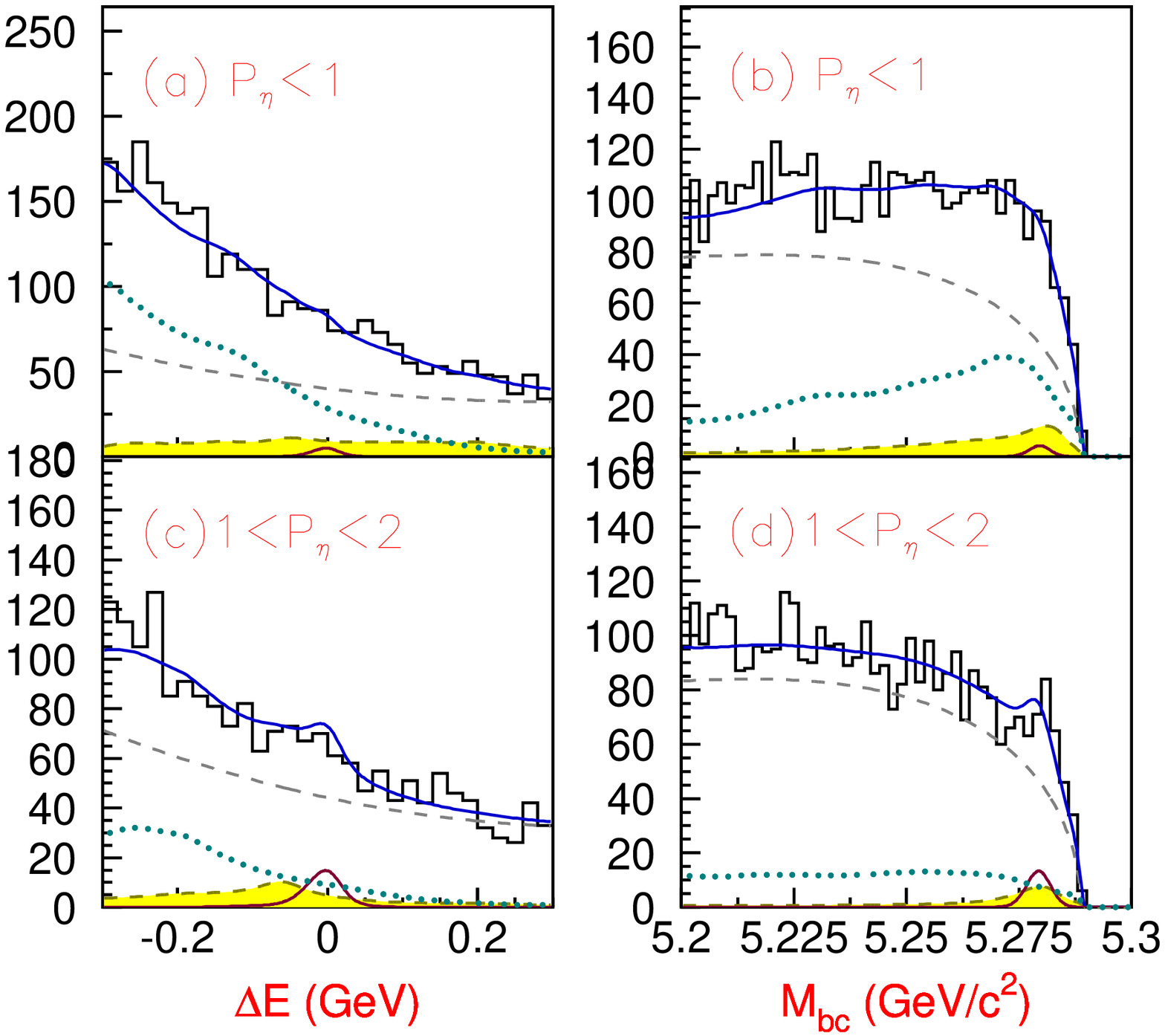}
\includegraphics[width=0.49\textwidth]{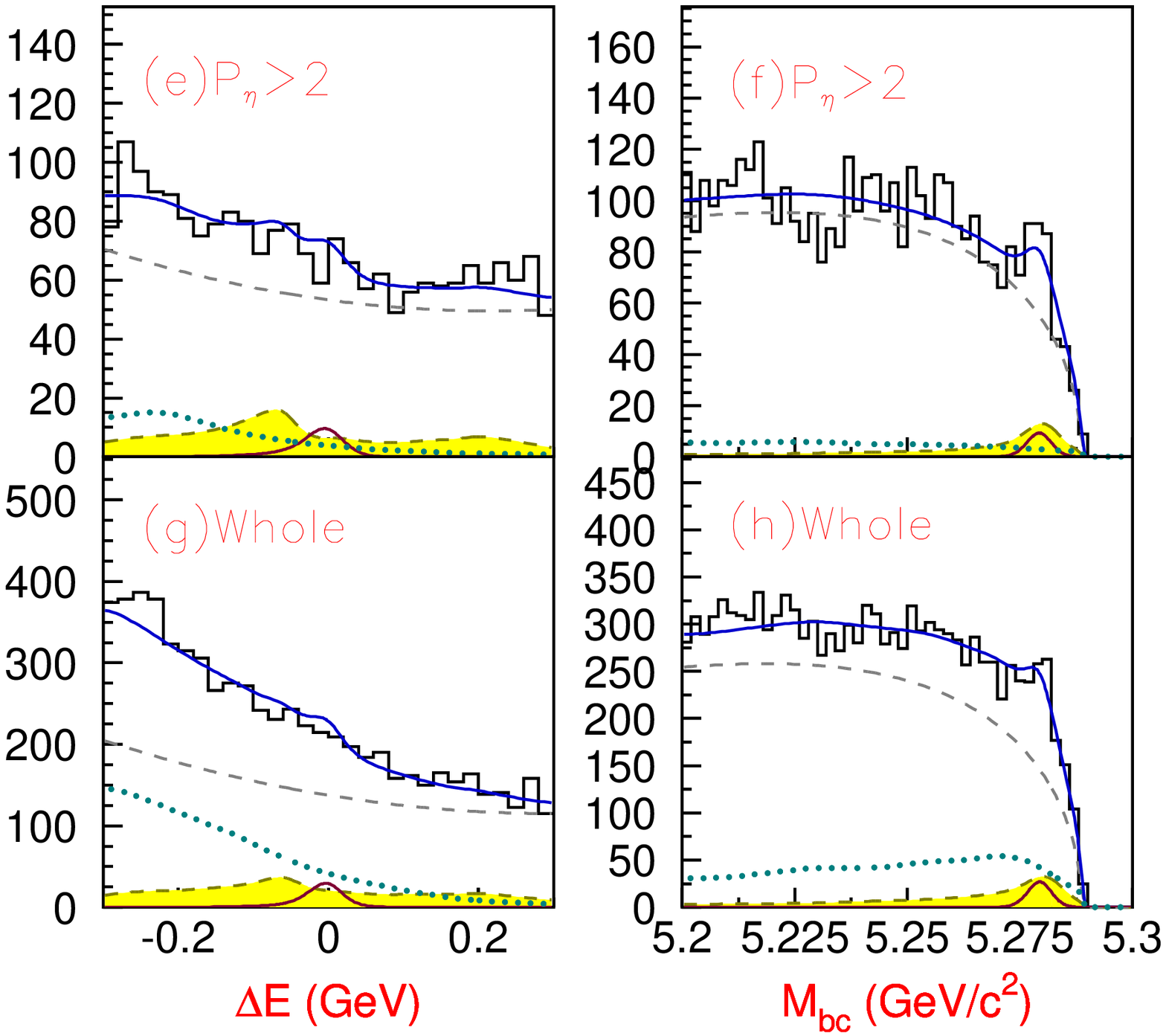}
\caption{\label{fitetapipi}Projections onto $M_{\rm{bc}}$ and $\Delta E$ from the extended unbinned 2D ML fit for inclusive $B^0 \to \eta \pi^+ \pi^-$ . The fit results with different $\eta$ momenta ranges are shown in the different plots. (a),(b) : $P_{\eta}<$ 1 GeV/c, (c),(d):
1 GeV/c $<$ $P_{\eta}$ $<$ 2 GeV/c, (e),(f) : $P_{\eta}>$ 2 GeV/c, (g),(h) : whole $P_\eta$ region. Components are the same as in Figure 1.}
\end{figure}

Signal yields obtained from the two dimensional
$M_{\rm{bc}}$ and $\Delta E$ extended unbinned maximum likelihood fits are shown in Table~\ref{fityield-3body}. The backgrounds from generic $B\overline{B}$ decays and other rare B backgrounds are considered in the fit. For the mode $B^0 \to \eta K^+ \pi^-$, the largest component of the rare decay background is $B \to \eta'K$. For the sample of $B^0 \to \eta \pi^+ \pi^-$, most of the rare B events come from $B^-\to \eta \pi^-$, $B^- \to \rho^- \eta$ and $B^0 \to \eta K^{*0}$. We use different 2D smooth functions as signal PDFs in the different $P_{\eta}$ regions. Fig.~\ref{fitetakpi} and Fig.~\ref{fitetapipi} show the $\Delta E$ and $M_{\rm{bc}}$ projections for the entire $\eta$ momentum range and the three sub-ranges. The fit results are summarized in Table ~\ref{fityield-3body}.

The yield of $B^0 \to K^* \eta$ accounts for about half of 
 the inclusive $B^0 \to \eta K^+\pi^-$ yield. The other half is not well understood so the default signal efficiency, $\epsilon_{sig}$, is based on a model with 75\% $B^0 \to K^*\eta$ and 25\% phase space. The difference between $\epsilon_{sig}$ for this model  and a model with 50\% $B^0 \to K^* \eta$ and 50\% phase space is included in the systematic error. In the case of $B^0 \to \eta\pi^+\pi^-$, there is no obvious enhancement in the two body mass spectra and the yields in different $\eta$ momentum ranges are similar to the expectation from phase space $B^0 \to \eta \pi^+ \pi^-$. Therefore we use a phase space model of $B^0 \to \eta \pi^+ \pi^-$ to determine $\epsilon_{sig}$ for inclusive $B^0 \to \eta \pi^+ \pi^-$.
\begin{table}[t]
\begin{center}
\caption{The fit yields from the two dimensional
$M_{\rm{bc}}$ and $\Delta E$ extended unbinned maximum likelihood fit for inclusive $B^0 \to \eta
K^+ \pi^-$ and $B^0 \to \eta \pi^+ \pi^-$.}
\vspace{0.4cm}
\label{fityield-3body}
\begin{tabular}{c|cccccc}
\hline
\hline
$\eta$ momentum(GeV/c) & $P_{\eta}$ $<$ 1 & 1 $<$ $P_{\eta}$ $<$ 2 & $P_{\eta}$ $>$ 2 & whole \cr
\hline
$B^0 \to \eta K^+ \pi^-$, $\eta \to \gamma \gamma$ & 25.9 $^{+13.4}_{-12.2}$ & 122.9 $^{+21.0}_{-19.7}$ & 409.7 $\pm$ 28.7 & 558.6 $^{+38.0}_{-36.9}$ \cr
$B^0 \to \eta K^+ \pi^-$, $\eta \to 3\pi$ & 3.8 $\pm$ 7.7 & 26.6 $^{+9.9}_{-8.6}$ & 111.4 $^{+14.5}_{-13.5}$ & 141.8 $^{+19.1}_{-17.7}$ \cr
$B^0 \to \eta \pi^+ \pi^-$, $\eta \to \gamma \gamma$ & 7.0 $^{+10.8}_{-9.4}$ & 44.3 $^{+13.9}_{-12.6}$ & 32.0 $^{+15.7}_{-14.5}$ & 83.3 $^{+23.6}_{-21.4}$ \cr
$B^0 \to \eta \pi^+ \pi^-$, $\eta \to 3\pi$ & 11.1 $^{+8.6}_{-7.5}$ & -3.0 $^{+8.1}_{-6.7}$ & 4.4 $^{+7.7}_{-6.4}$& 12.4 $^{+14.0}_{-11.8}$ \cr
\hline
\hline
\end{tabular}
\end{center}
\end{table}

\begin{figure}[htb]
\includegraphics[width=0.45\textwidth]{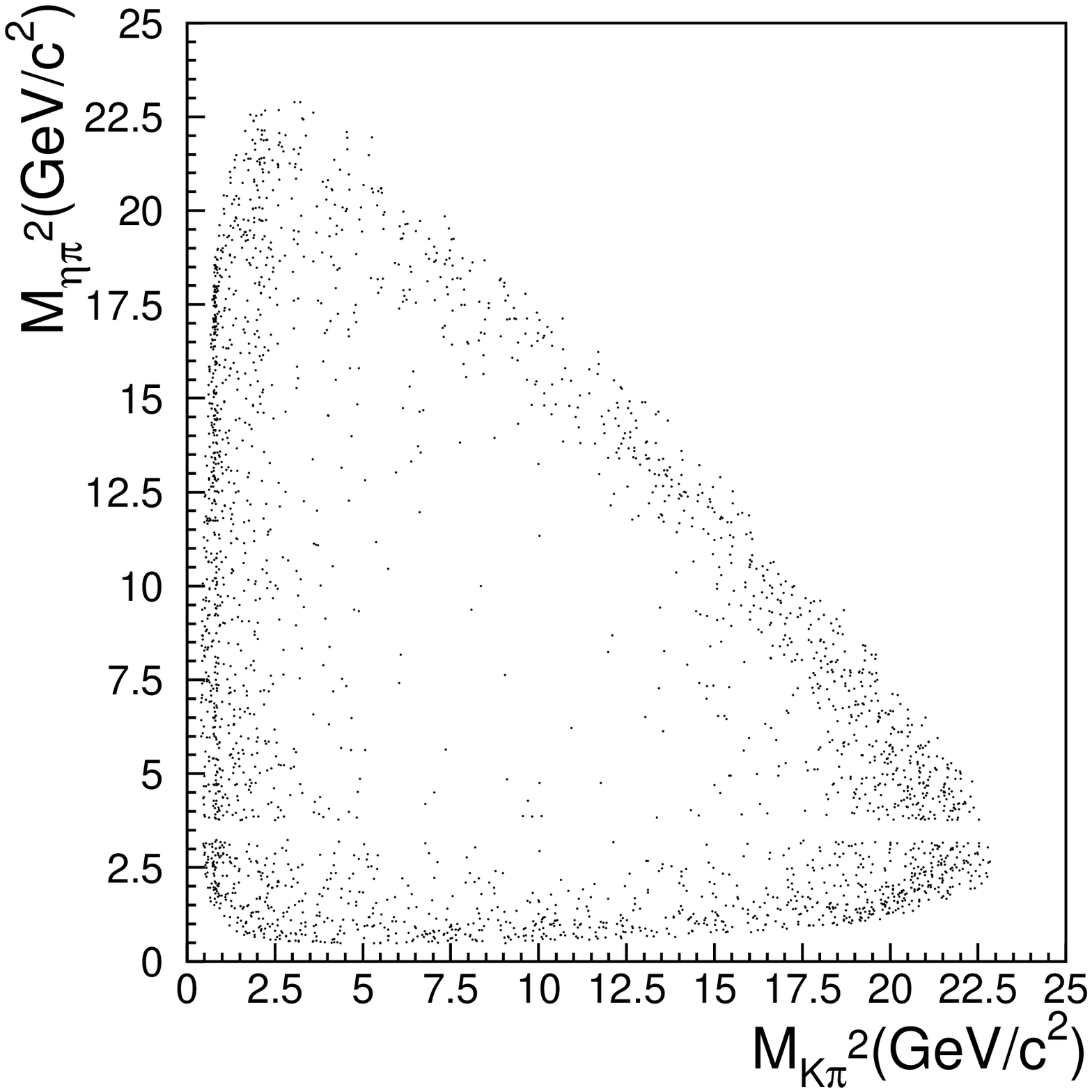}
\includegraphics[width=0.45\textwidth]{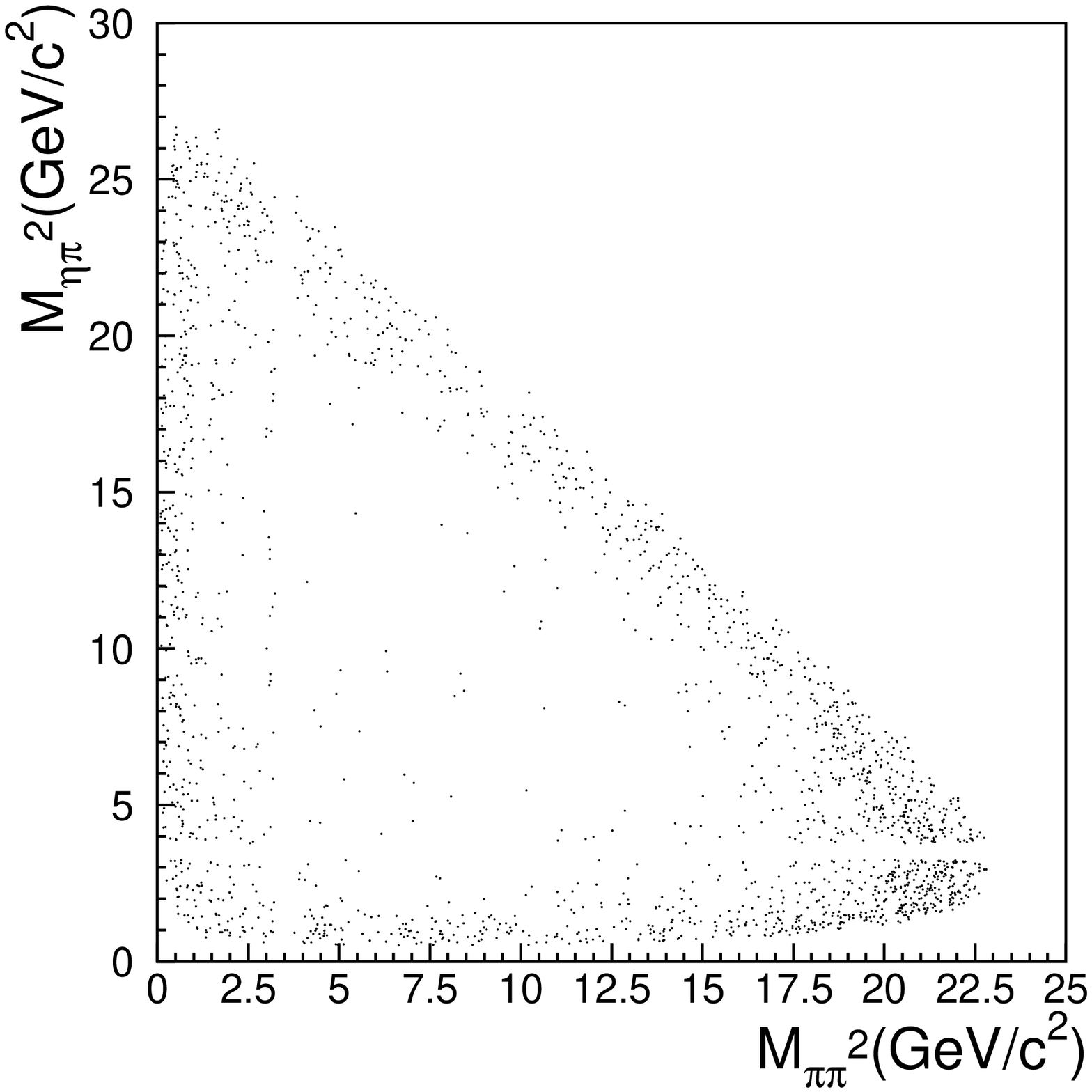}
\caption{\label{dalitz}Dalitz plots of (a)$B^0 \to \eta K^+ \pi^-$ and (b)$B^0 \to \eta \pi^+ \pi^-$ candidates from the $B$ signal region. We have applied the $D$ veto in both plots.}
\end{figure}
 
\subsection{Two body mass spectra of $\eta K^+ \pi^-$  and $\eta \pi^+ \pi^-$ in the final state}
\begin{figure}[htb]
\includegraphics[width=0.43\textwidth]{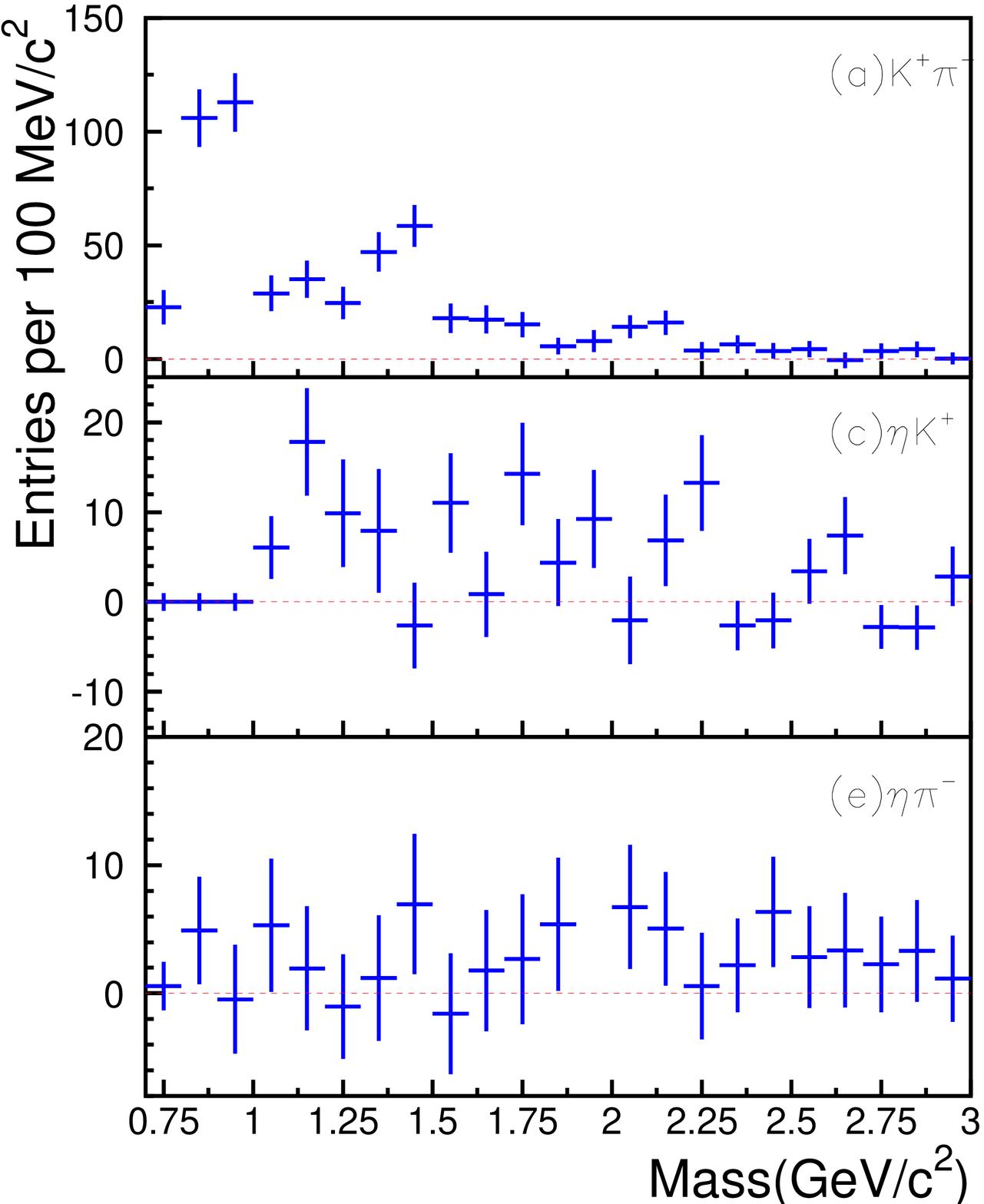}
\includegraphics[width=0.43\textwidth]{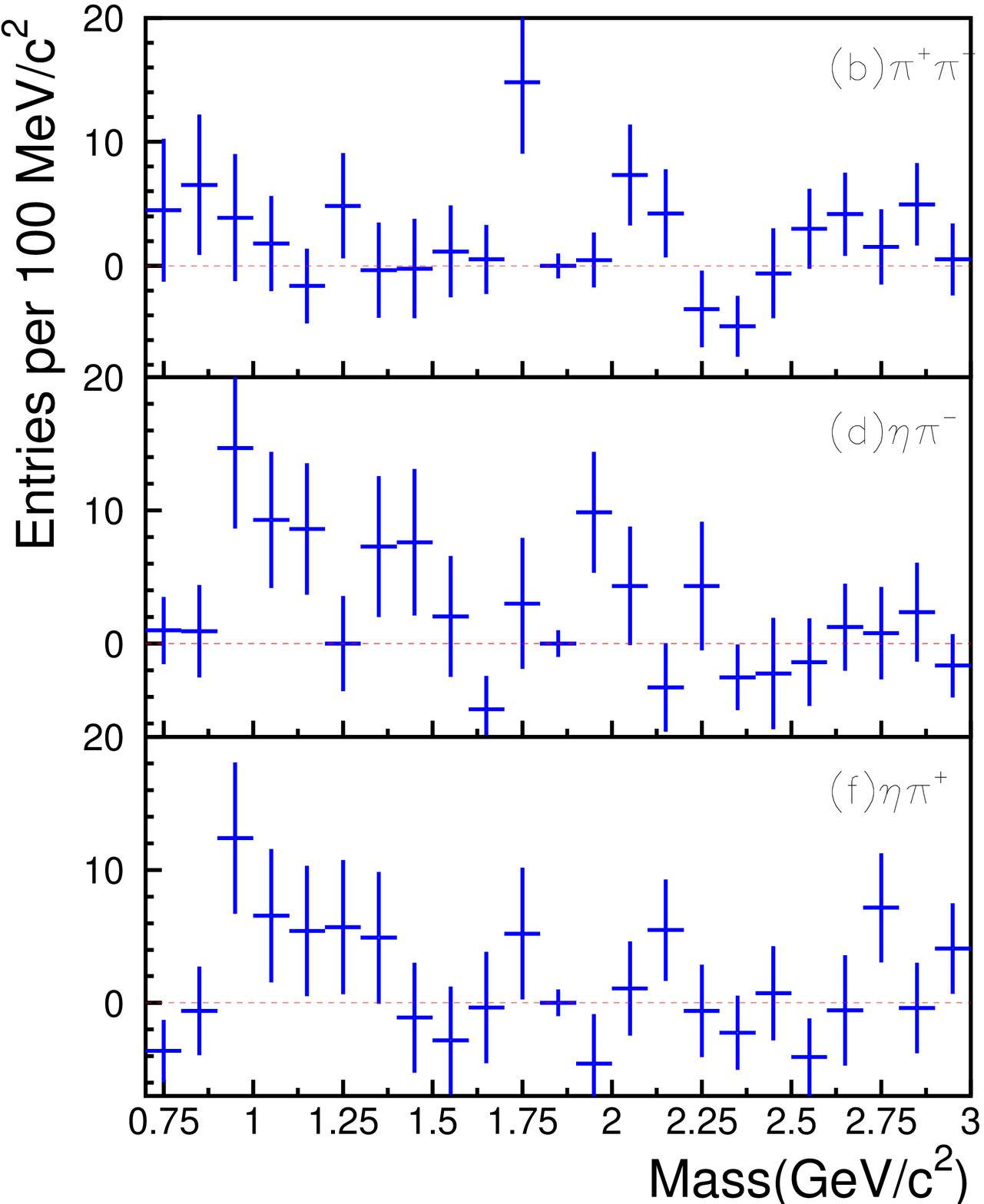}
\caption{\label{2bodyetakpi}The left hand plots are signal yields from the 2-D fit as a function of (a)$M_{K^+ \pi^-}$, (c)$M_{K^+ \eta}$ and (e)$M_{\pi^- \eta}$ with the $B^0 \to \eta K^+ \pi^-$ final state. The right hand plots are signal yields from the 2-D fit as a function of (b)$M_{\pi^+ \pi^-}$, (d)$M_{\eta \pi^-}$ and (f)$M_{\eta \pi^+}$ with the $B^0 \to \eta \pi^+ \pi^-$ final state.}
\end{figure}

The Dalitz distributions are densely populated close to the edges, as can be seen in Fig.~\ref{dalitz}, indicating possible resonances in our sample. We investigate
these by examining the 1-D projections of the Dalitz plot.
We divide each
2-body mass distribution into 100 MeV/$c^2$ bins. A 2-D fit is then applied in each bin 
 to determine the $B$ yield. To prevent cross talk from resonances in the $K\pi$ region, we require the 
$K\pi$ mass to be larger than 2 GeV/c$^2$ when showing the $K\eta$ and $\pi \eta$ mass distributions~\cite{kpipi0}.
For the $B^0 \to \eta K^+ \pi^-$ final state, the signal PDF is based on a mixture of $B^0\to \eta K^{*0}$ and $B^0 \to \eta K^+ \pi^-$ MC samples.
Fig.~\ref{2bodyetakpi} shows the signal yields obtained from the 2-D fits as functions of the 2-body masses. There is an obvious enhancement from the $K^{*0}$(892) resonance and an excess in the $K \pi$ mass region between 1.4 GeV/c$^2$ and 1.7 GeV/c$^2$ which may be due to $K_0^*$(1430) or $K_2^*$(1430). For $B^0 \to \eta \pi^+ \pi^-$, the signal PDF is based on phase space MC. Fig.~\ref{2bodyetakpi} shows the results. In
the $\pi^+ \pi^-$ mass distribution, there is a small enhancement at low mass, which may be due to either the $\rho^0$ or $f_0$(980).
In the $\eta \pi^{\pm}$ mass distribution, there is a small excess in the 1 GeV/$c^2$ region that may be from the $a_0^{\pm}$. 
This is discussed in greater detail in the next section.
\subsection{$B^0 \to a_0^- K^+$ and $B^0 \to a_0^- \pi^+$}
\begin{figure}[htpb]
\includegraphics[width=0.7\textwidth]{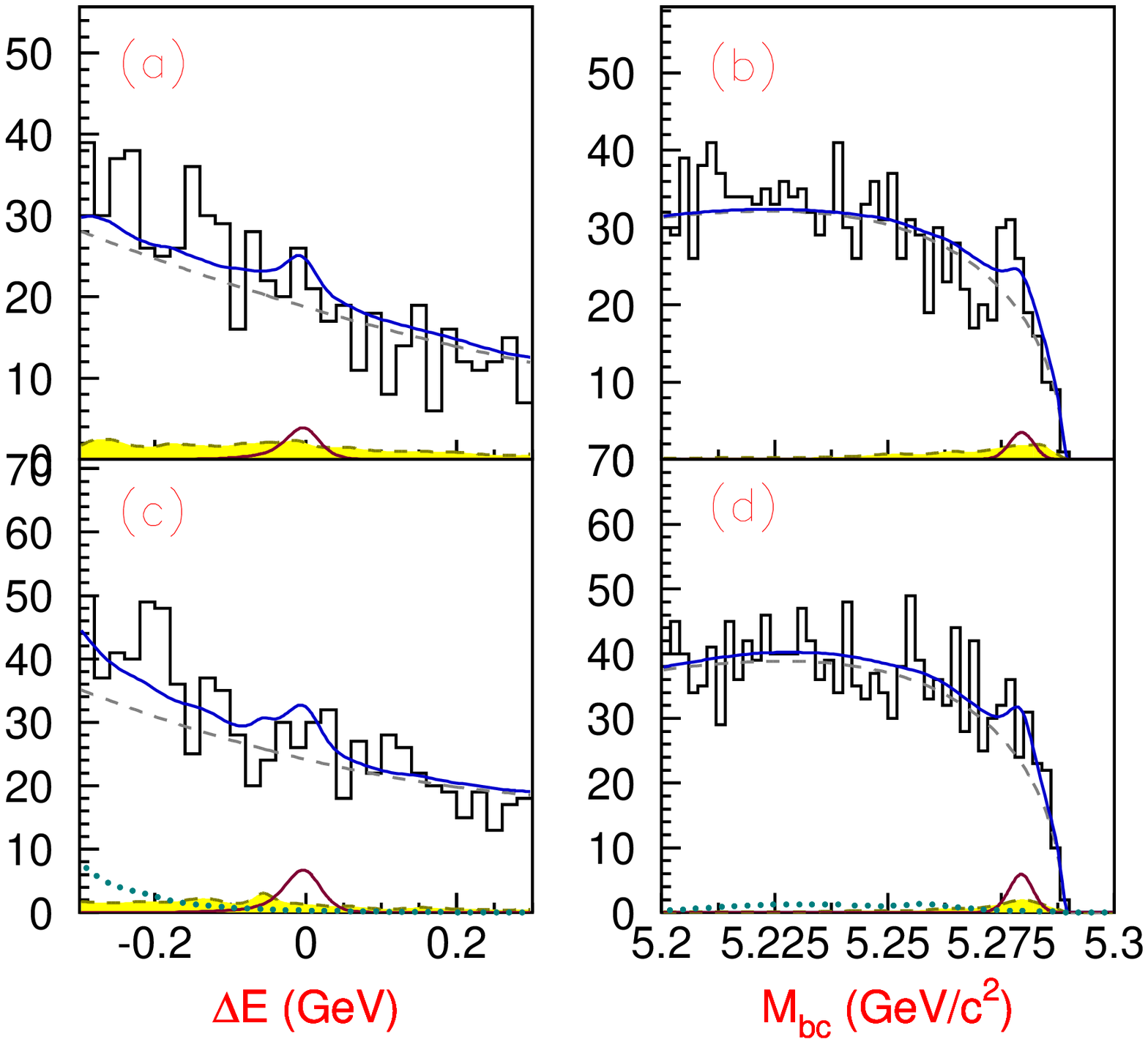}
\caption{\label{total} Projections onto $M_{\rm{bc}}$ and $\Delta E$ from the extended unbinned 2D ML fits
for $B^0 \to a_0^- K^+$ are shown in (a) and (b). The projectoions for $B^0 \to a_0^- \pi^+$ are shown in (c) and (d). Components are the same as in Figure 1.}
\end{figure}

In the $B^0 \to a_0^-K^+$, $a_0^- \to \eta \pi^-$ mode, the background from other rare $B$ decay modes is included only in the $\eta \to \gamma \gamma$ subdecay mode, otherwise it is ignored due to the negligible contribution predicted from MC. For the
$\eta \to \pi^+ \pi^- \pi^0$ subdecay mode, the generic $B\overline{B}$ background is neglected for the
same reason.
Figure~\ref{total}(a) and \ref{total}(b) show the $M_{\rm{bc}}$ and $\Delta E$ projections from the extended unbinned 2D ML fit for $B^0 \to \eta K^+\pi^- $, with $M_{\eta \pi^-}$ in the $a_0^-$ mass region. From the fit, the yields for the $\eta \to \gamma \gamma$ and the $\eta \to \pi^+ \pi^- \pi^0$ subdecay modes are 17.3$^{+9.2}_{-8.0}$ and -4.7 $^{+3.8}_{-2.5}$, respectively. These yields are the sum of $B^0 \to a_0^- K^+$ and non-resonant $B^0 \to \eta K^+\pi^-$.

In the $B^0 \to a_0^- \pi^+$, $a_0^- \to \eta \pi^-$, $\eta \to \gamma \gamma$ mode, we use smoothed 2-D histograms to model the $B\overline{B}$ and rare $B$ backgrounds. The fractions of these backgrounds are fixed from the MC. In the $\eta \to \pi^+ \pi^- \pi^0$ mode, we consider only the $b$ $\to$ $c$ background and neglect backgrounds from other rare $B$ decays. Figure~\ref{total}(c) and \ref{total}(d) show the $M_{\rm{bc}}$ and $\Delta E$ projections from extended unbinned the 2D ML fit for $B^0 \to \eta \pi^+ \pi^-$, with $M_{\eta \pi^-}$ in the $a_0^-$ mass region. From the fit, the yields for the $\eta \to \gamma \gamma$ and the $\eta \to \pi^+ \pi^- \pi^0$ subdecay modes are 22.6 $^{+10.1}_{-8.9}$ and -0.4 $^{+5.1}_{-3.8}$, respectively. Again note that the yields are the sum of $B^0 \to a_0^- \pi^+$ and non-resonant $B^0 \to \eta \pi^+ \pi^-$.
\begin{table}[hb]
\caption{Summary of systematic errors for inclusive $B^0 \to \eta K^+\pi^- $ and inclusive $B^0 \to \eta \pi^+ \pi^-$.
The systematics from $\eta$ and $\pi^0$ reconstruction are $\sigma_{\eta}$ and
$\sigma_{\pi^0}$ respectively; $\sigma_{pid}$ is the systematic from PID;
$\sigma_{N_{B\bar B}}$ is the systematic from the number of BB pairs in our data;
$\sigma_{sig}$ is the systematic due to uncertainty in the signal PDF;
$\sigma_{fit}$ contains systematics from other fit component PDFs;
$\sigma_{tr}$ is the systematic from charged tracking; and
$\sigma_{R}$ is from the $\mathcal R$ cut.
All are expressed as percentages (\%).}
\begin{tabular}{cccccccccc} \hline
decay mode & $\sigma_{tr}$ & $\sigma_{R}$ & $\sigma_{\eta}$ & $\sigma_{\pi^0}$ & $\sigma_{pid}$ & $\sigma_{N_{B\bar B}}$ & $\sigma_{sig}$ & $\sigma_{fit}$ & Sum \cr
\hline
$B^0 \to \eta_{\gamma \gamma} K^+\pi^- $ & 2.0 & 2.1 & 3.0 & 0.0 & 1.5 & 1.0 & 4.0 & $^{+3.3}_{-4.6}$ & $^{+6.9}_{-7.6}$ \cr
$B^0 \to \eta_{\pi \pi \pi^0} K^+\pi^- $ & 4.1 & 2.1 & 3.0 & 3.0 & 2.5 & 1.0 & 4.2 & $^{+4.8}_{-8.2}$ & $^{+9.4}_{-11.5}$ \cr
\hline
$B^0 \to \eta K^+\pi^- $ & 2.5 & 2.1 & 3.0 & 0.3 & 1.6 & 1.0 & 4.1 & $^{+3.4}_{-5.4}$ & $^{+7.1}_{-8.2}$ \cr
\hline
$B^0 \to \eta_{\gamma \gamma} \pi^+ \pi^-$ & 2.0 & 2.0 & 3.0 & 0.0 & 1.2 & 1.0 & 0.0 & $^{+5.2}_{-4.8}$ & $^{+6.8}_{-6.5}$ \cr
$B^0 \to \eta_{\pi \pi \pi^0} \pi^+ \pi^-$ & 4.2 & 2.0 & 3.0 & 3.0 & 2.5 & 1.0 & 0.0 & $^{+18.6}_{-13.2}$ & $^{+19.8}_{-14.8}$ \cr
\hline
$B^0 \to \eta \pi^+ \pi^-$ & 2.6 & 2.0 & 3.0 & 0.9 & 1.4 & 1.0 & 0.0 & $^{+11.6}_{-8.3}$ & $^{+12.5}_{-9.6}$ \cr
\hline
\hline
\end{tabular}
\label{sys}
\end{table}
 
\begin{table}[thb]
\begin{center}
\caption{
Fitting significance($\Sigma$) and
branching fractions(${\mathcal B}$)
from the extended unbinned $\Delta E$-$M_{\rm{bc}}$ 2-D ML fits. The significance of $a_0^- X^+$ includes possible non-resonant events in the selected mass region.}
\begin{tabular}{lcccccc} \hline\hline
Mode &  $\Sigma$  & $\mathcal B$($10^{-6}$)  \cr
\hline \hline
inclusive $B^0 \to \eta K^+ \pi^-$ & 24.0 & 31.7 $\pm$ 1.9 $^{+2.2}_{-2.6}$\cr
\hline
$B^0 \to a_0^- K^+$ & 2.0 & ($<$ 1.6) \cr
\hline
inclusive $B^0 \to \eta \pi^+ \pi^-$ & 3.6 & 6.2 $^{+1.8+0.8}_{-1.6-0.6}$($<$11.9) \cr
\hline
$B^0 \to a_0^-\pi^+$ & 1.4 &($<$2.8)\cr
\hline\hline
\end{tabular}
\label{result}
\end{center}
\end{table}

\section{Systematic Error}
Systematic errors may arise from
the efficiency corrections and from the fitting process.
The main sources of uncertainties in the efficiency
corrections are from
the reconstruction of low-momentum charged tracks,
low-energy photon finding, and ${\mathcal R}$ cut efficiency,
each at the level of a few percent.
The systematic errors include contributions of
$2$\% for the ${\mathcal R}$ cut, $1$\% per reconstructed charged particle for tracking, $0.5$\%
per charged particle for PID,
$3$\% for $\pi^0$ reconstruction,
and $3$\% for $\eta$ reconstruction. 
The fitting systematic errors are estimated by varying the fitting function PDF variables by $\pm$ 1 $\sigma$ from the measured values. The variations in the fitted yields are then quadratically summed
to get the total fit systematic uncertainty.
The contributions to the systematic error are given in Table~\ref{sys}. 

The intrinsic width of $a_0^-$ is 57 MeV/c$^2$ for the MC samples of $B^0 \to a_0^-K^+$ and $B^0 \to a_0^-\pi^+$ used to calculate the efficiencies.
If instead we assume the intrinsic width of $a_0^-$ to be 100 MeV/c$^2$~\cite{pdg}, the signal efficiency decreases by 9.7\%.
This uncertainty is considered in the upper limit calculation.

\section{Discussion and Conclusion}
 
In summary, we have searched for inclusive charmless hadronic
$B^0 \to \eta K^+\pi^-$ and $B^0 \to \eta \pi^+ \pi^-$ decays and have observed
the branching fractions listed below, we also give upper limits at 90\% C.L. for the inclusive $B^0 \to \eta \pi^+ \pi^-$ :
 
\begin{eqnarray*}
{\mathcal B}(B\to\eta K^+\pi^-)&=&
(31.7 \pm 1.9 ^{+2.2}_{-2.6})\times 10^{-6}, \\
{\mathcal B}(B\to\eta \pi^+ \pi^-)&=&
(6.2^{+1.8+0.8}_{-1.6-0.6})\times 10^{-6}(<11.9 \times 10^{-6}). \\
\end{eqnarray*}
 
We do not find significant signals for $B^0 \to a_0^-K^+$ or $B^0 \to a_0^- \pi^+$.
The 90\% C.L. upper limits on the respective branching fractions are:
\begin{eqnarray*}
{\mathcal B}(B^0 \to a_0^- K^+)
<1.6 \times 10^{-6}, \\
{\mathcal B}(B^0 \to a_0^- \pi^+)
<2.8 \times 10^{-6}.
\end{eqnarray*}
The notation $\mathcal B$($B^0 \to a_0^- X^+$) indicates the product of branching fractions for $B^0 \to a_0^- X^+$ and $a_0^- \to \eta \pi^-$, where $X^+$ is a $K^+$ or $\pi^+$.
All the results are listed in Table~\ref{result}.
 
 \section{Acknowledgments}
We thank the KEKB group for the excellent operation of the
accelerator, the KEK cryogenics group for the efficient
operation of the solenoid, and the KEK computer group and
the National Institute of Informatics for valuable computing
and Super-SINET network support. We acknowledge support from
the Ministry of Education, Culture, Sports, Science, and
Technology of Japan and the Japan Society for the Promotion
of Science; the Australian Research Council and the
Australian Department of Education, Science and Training;
the National Science Foundation of China under contract
No.~10175071; the Department of Science and Technology of
India; the BK21 program of the Ministry of Education of
Korea and the CHEP SRC program of the Korea Science and
Engineering Foundation; the Polish State Committee for
Scientific Research under contract No.~2P03B 01324; the
Ministry of Science and Technology of the Russian
Federation; the Ministry of Higher Education,
Science and Technology of the Republic of Slovenia;
the Swiss National Science Foundation; the National Science Council and
the Ministry of Education of Taiwan; and the U.S.\
Department of Energy.

\end{document}